# $J_c$ Scaling and Anisotropies in Co-doped Ba-122 thin films

Jens Hänisch, Kazumasa Iida, Silvia Haindl, Fritz Kurth, Alexander Kauffmann, Martin Kidszun, Thomas Thersleff, Jens Freudenberger, Ludwig Schultz, and Bernhard Holzapfel

*Abstract*—We have successfully grown epitaxial, superconducting films of Ba(Fe$_{1-x}$Co$_x$)$_2$As$_2$ (Ba-122) with $x \sim 0.1$. The films grow without observable correlated defects parallel to the c-axis, as confirmed by TEM. This is also reflected in the absence of a c-axis peak in $J_c(\theta)$. In contrast to cuprate high-$T_c$ superconductors such as YBa$_2$Cu$_3$O$_{7-\delta}$ or even Bi$_2$Sr$_2$Ca$_2$Cu$_3$O$_{10-\delta}$, the pnictides show a rather low anisotropic behavior in their $J_c(\theta)$ behavior as well as in their upper critical fields, $H_{c2}$. As a multiband superconductor, Ba-122 exhibits a temperature dependent electronic mass anisotropy.

## I. INTRODUCTION

THE recent discovery of high-temperature superconductivity in Fe-As containing oxides [1] and intermetallics [2] sparked world-wide research activities to investigate the fundamental properties of these extraordinary materials, to look for possible higher transition temperatures and even to investigate application possibilities [3].

The pnictides show a very interesting combination of properties, spanning the range between MgB$_2$ and the cuprates. The maximal critical temperature measured so far lies between these two classes of superconductors (Sm(O$_{1-x}$F$_x$)FeAs: ~55 K [4]). The pnictides have a layered crystal structure like the cuprates and equally short coherence lengths, leading to similar grain boundary effects [5], [6]. They are multiband superconductors [7] as MgB$_2$, and – depending on doping level and homogeneity – may or may not show Pauli limiting of the upper critical field [5], [8].

Early investigations of the pnictide superconductors were performed on polycrystalline bulk samples [9]. However, to determine the anisotropy of physical properties, single-crystalline samples or epitaxial thin films are needed [10]. Thin films may have different advantages, especially if single crystals are not available (yet) or extremely difficult to produce. First, the reduced dimensions allow the measurement of geometry effects, especially if the thickness is smaller or comparable to some critical dimension, such as coherence length or London penetration depth. Second, thin films facilitate transport measurements of the critical current density, in contrast to single crystals where in the case of high critical currents the sample is heated up. Third, thin film techniques enable the preparation of structures not easily possible in single-crystal growth, such as multilayers and metastable phases. Finally, thin films have a wide potential for applications of superconductors, as seen in electronics and coated conductors of cuprate high-$T_c$ superconductors.

In one-band superconductors, which are not in the two-dimensional limit and whose upper critical field is not Pauli-limited, the electronic mass anisotropy, $\gamma_m$, determines both the anisotropy of the penetration depth, $\gamma_\lambda$ (and hence of the lower critical field $H_{c1}$), and the anisotropy of the coherence length, $\gamma_\xi$ (and hence the upper critical field, $\gamma_H$). In multiband superconductors, these anisotropies are in general not equal, and furthermore, temperature dependent. It is not a priori obvious which behavior $\gamma_m$ will show and which of the other anisotropies it will follow. To illuminate this question, we investigate the scaling behavior of the critical current density $J_c$ in Ba(Fe$_{0.9}$Co$_{0.1}$)$_2$As$_2$ (Ba-122) thin films of high crystalline quality. After a short overview of the deposition process and the basic structural and superconductive properties, different types of anisotropies in this material are compared and their temperature dependencies discussed with respect to the two-band nature of superconductivity in this material.

## II. EXPERIMENTAL

### A. Film Preparation

The films are prepared by pulsed laser deposition (PLD) using a KrF excimer laser ($\lambda$ = 248 nm) with an energy density of around 4 J/cm$^2$ at the target surface. The target was prepared by pressing a stoichiometric mixture of Ba$_5$As$_3$, Fe$_2$As, FeAs, and Co into a pellet and annealing it for 16 h at 900 °C in an evacuated quartz tube.

Recently, we have shown that the choice of the substrate strongly influences the critical temperature of the film due to the interfacial strain in very thin films [11]. Although SrTiO$_3$ (STO) substrates resulted in the highest $T_c$ values reported so far (24.5 K), we chose (001) (La,Sr)(Ta,Al)O$_3$ (LSAT) because STO becomes conducting at the deposition temperature of 700 °C under UHV condition (base pressure 10$^{-9}$ mbar) due

Manuscript received 3 August 2010.
This work was supported in part by the German Research Foundation (DFG) under project SPP 1458.
All authors are with the Institute for Metallic Materials at IFW Dresden, Helmholtzstr. 20, 01069 Dresden, Germany (corresponding author: J. Hänisch, phone: +49-351-4659-607; fax: +49-351-4659-9607; e-mail: j.haenisch@ifw-dresden.de). J. Hänisch is currently with Superconductivity Technology Center at Los Alamos National Laboratory, Los Alamos, NM 87545, USA.

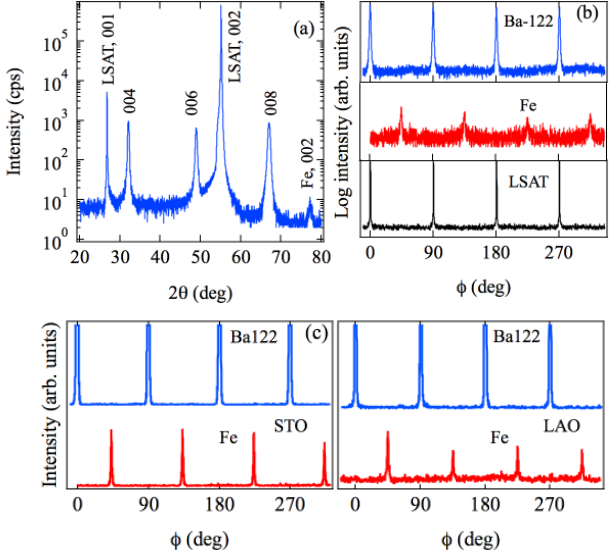

Fig. 1. a) XRD scans (CoK$_\alpha$) in Bragg-Brentano geometry for a Ba-122 thin film on LSAT. The film shows excellent c-axis textured growth and no signs for impurity phases in the superconducting layer. b) Φ-scans of LSAT substrate (110 reflection), Fe interlayer (110), and Ba-122 (103), showing perfect epitaxial growth. c) Φ-scans of Ba-122 and naturally grown Fe interlayer on two different substrates, LaAlO$_3$ and SrTiO$_3$.

to the loss of oxygen. The laser repetition rate was 5 Hz, and for protection against ambient conditions, a thin Cr layer was deposited on top [12]. The critical temperature, $T_c$, and the upper critical field, $H_{c2}$, were measured resistively and defined as 90% of the normal resistance just above $T_c$. The Ba-122 film had a $T_c$ of 22.4 K. For transport measurements in a Quantum Design PPMS for static fields up to 9 T and in a hybrid magnet for pulsed fields up to 45 T [15], thin bridges of 1-2 mm width and 10 mm length were cut mechanically. Electrical contacts were made by Cu wires and silver paste. A four-probe technique was used with voltage taps separated by around 1 mm. The critical current density, $J_c$, was defined at an electrical field criterion of 1 µV/cm on $E(J)$ curves measured with current pulses of 30 ns duration.

### B. Structural Properties

Figs. 1a and 1b show the x-ray diffraction pattern for the Ba-122 film used in this study. No misorientations or impurity phases, except a (002) reflection of Fe are observed. Confirmed by TEM investigations [12], an epitaxial Fe layer is formed at the interface between substrate and Ba-122 film, most likely due to a strong As evaporation at the initial stage of film growth hindering the immediate Ba-122 phase formation. This Fe layer was observed in several films on different substrates (Fig. 1b, c). Recently, we were able to demonstrate that Fe can be used as artificial buffer layer for Ba-122 thin films [13], [14]. The films grow epitaxially with the relation (001)[100]Ba-122∥(001)[110]Fe∥(001)[100]LSAT. The thickness of the superconducting layer was measured to be 33 nm.

### III. ELECTRICAL TRANSPORT MEASUREMENTS

Figs. 2 and 3 show the temperature and the field dependence of the resistive transitions of the thin film, measured with an applied current of 100 µA and 95 µA, respectively. Already here the anisotropic behavior is evident, the transitions for **H**∥**c** cover a larger temperature range and a smaller field range than for **H**⊥**c**. The dashed lines in Fig. 3 are test measurements with different peak fields and currents to assure that sample-heating effects can be excluded.

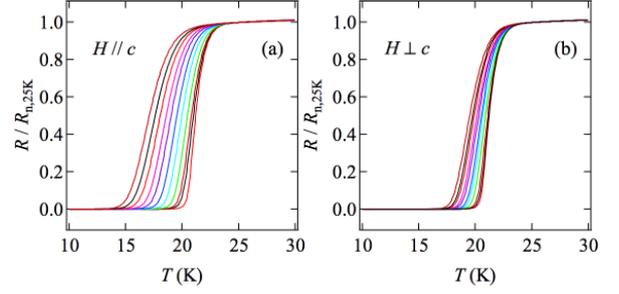

Fig. 2. $R(T)$ data measured in static magnetic fields in a PPMS with applied current of 100 µA For both major directions, **H**∥**c** (left) and **H**⊥**c** (right), normalized to $R(25\,K)$. Step size 1 T.

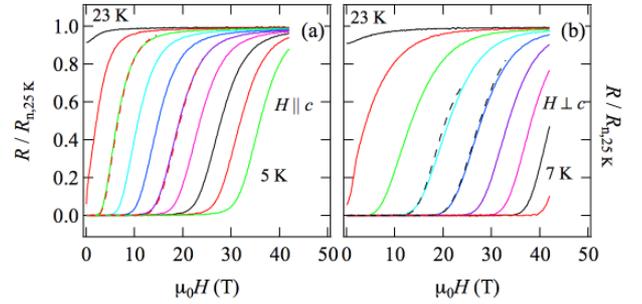

Fig. 3. $R(\mu_0 H)$ measured in a pulse field system with applied current of 95 µA For both major directions, **H**∥**c** (left) and **H**⊥**c** (right), normalized to $R(25\,K)$. The dashed lines are test measurements with different peak fields and current to exclude heating effects. Step size 2 K.

The angular dependence of $J_c$ at 7.5 K and 15 K and several magnetic fields is given in Fig. 4. The sample, with a self-field $J_c$ of 50 kA/cm$^2$ at 4.2 K, shows a clear $J_c$ anisotropy with a maximum at 90° (**H**⊥**c**). The layered crystal structure and the short coherence length parallel to the c-axis are responsible for a modulation of the order parameter in this direction which causes "intrinsic" pinning of flux lines in between the FeAs planes. This is very similar to cuprates, like $YBa_2Cu_3O_{7-\delta}$ (YBCO) or even $Bi_2Sr_2Ca_2Cu_3O_{10-\delta}$, (Bi2223) [16]. In contrast to many cuprate thin films and some pnictide films [17], our samples do not show a second peak at $\theta = 0°$ (**H**||**c**) at any applied field. This, together with the overall low $J_c$ values compared to cuprates, is explained by the absence of pinning-effective correlated defect structures in this direction, confirmed by TEM investigations [12], [13].

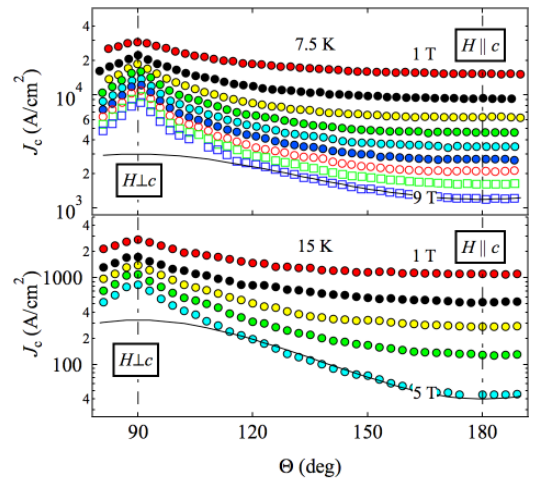

Fig. 4. Angular dependence of the critical current density $J_c$ for Ba-122 at 7.5 K and 15 K for several magnetic fields, step size 1 T. Crystalline, electronic mass anisotropy, and intrinsic pinning are responsible for the $J_c(\theta)$ variation with a maximum for **H**⊥**c** (90°). The lines represent the contribution of random pinning.

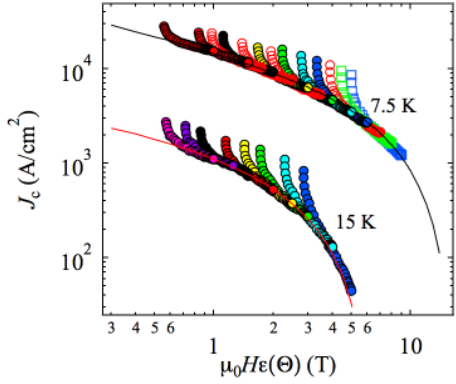

Fig. 5. Anisotropic Ginzburg-Landau scaling (1) at 7.5 K ($\gamma_m$ = 1.55) and 15 K ($\gamma_m$ = 1.7) for the data of Fig. 3 and some additional lower fields. In a wide angular range around **H**||**c** the data scale, only random defects are present for these angles. The solid lines are fits after the Kramer model and used for the random-pinning contribution of $J_c(\theta)$ in Fig. 4. For details see [24].

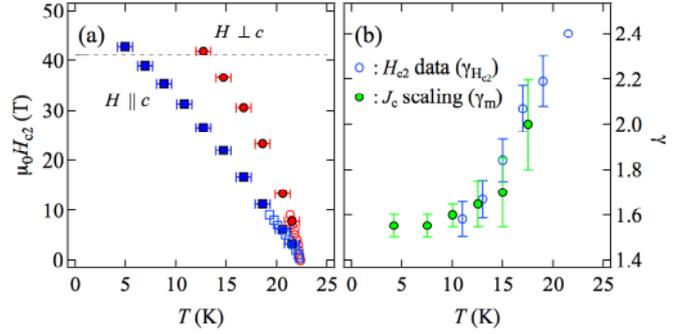

Fig. 6. a) Temperature dependence of the upper critical field $H_{c2}$ in both major directions, **H**||**c** (squares) and **H**⊥**c** (circles), open symbols measured statically in a PPMS, full symbols measured in pulsed fields up to 45 T. b) Temperature dependence of the mass anisotropy $\gamma_m$ (●) and the $H_{c2}$ anisotropy $\gamma_H$ (○) for Ba-122, The data point (21.5 K, 2.4) is the result of a full anisotropic-GL fit to $H_{c2}(\theta)$. Both anisotropies overlap and show a monotonous decrease with decreasing temperature.

## IV. DISCUSSION

In the frame of the anisotropic Ginzburg-Landau (AGL) theory with only randomly distributed and isotropically shaped pinning centers, i.e. atomic-scale lattice defects (e.g. vacancies) or nanoparticles, the $J_c$ anisotropy only depends on the electronic mass anisotropy of the participating charge carriers and, hence, $J_c(\theta)$ (Fig. 4) can be *scaled* with an effective magnetic field $H_{eff} = H \cdot \varepsilon$, where

$$\varepsilon = \sqrt{\cos^2(\theta) + \gamma_m^{-2} \sin^2(\theta)} \qquad (1)$$

represents the GL anisotropy and $\gamma_m = (m_c/m_{ab})^{1/2}$ the electronic mass anisotropy [18]. The scaling works best for higher fields, firstly because at low fields the applied field and the flux lines are not necessarily parallel to each other [19] (a prerequisite for anisotropic GL scaling), secondly because the $J_c$ range is larger, leading therefore to a higher accuracy of the scaling. The scaling procedure can provide information about different pinning regimes if $\gamma_m$ is known [20]. The angular range of $J_c$ where correlated defects dominate the pinning (e.g. the interlayers of reduced Cooper pair density in layered superconductors (intrinsic pinning) or dislocations and other one-dimensional defects, e.g. in c-direction) will not scale with (1), in contrast to the angular range where randomly distributed and isotropically acting defects are present. If, on the other hand, it is known that mainly the latter defects are present, the electronic mass anisotropy may be extracted, as shown for $BaZrO_3$-doped YBCO films with very low "effective" $\gamma_m$ values [21]. Even though GL theory is valid in the vicinity of $T_c$, it is usually applicable in a wide temperature range for high-$T_c$ superconductors because of the short coherence length.

Since our films do not show signs of c-axis correlated disorder, we may test the scaling behavior of the $J_c$ data around the c-direction. This is shown in Fig. 4 for the data of Fig. 5. $J_c(\theta)$ does indeed scale with (1) in a wide angular range around **H**||**c** for $\gamma_m$ = 1.55±0.05 at 7.5 K and 1.65 ± 0.2 at 15 K. The defects responsible for pinning are most likely point defects such as vacancies and disorder at the doping sites.

For single-band superconductors, the mass anisotropy is generally temperature independent. This is not necessarily the case for multi-band superconductors, where the bands can differ in critical temperatures, carrier types (electrons or holes) and effective masses. Indeed, applying the scaling at several temperatures for this sample requires significantly different values of $\gamma_m$ to obtain reasonable scaling. The results are depicted in Fig. 6b. The scaling parameter – here interpreted as a resulting effective mass anisotropy, $\gamma_m$, which is a combination of the mass anisotropies of the individual bands – is decreasing with decreasing temperature, spanning a range between 1.5 and 2.2.

Fig. 6a shows the temperature dependence of the upper critical field, $H_{c2}$, for both directions, **H**||**c** and **H**⊥**c**, the latter resulting in smaller values. The anisotropy ratio $\gamma_H = H_{c2}^{\perp c}/H_{c2}^{\parallel c}$ (Fig. 6b) is decreasing with decreasing temperature. Furthermore, $\gamma_m$ and $\gamma_H$ are overlapping in a large temperature range within the error bars, showing that using the AGL scaling of $J_c$, we seem to probe the anisotropy of the upper critical field. This is remarkable because in case of multiband superconductors, the angular dependence of the upper critical field cannot be described with (1) at all temperatures because it depends on the mass anisotropies and the diffusivities of the different bands, as well as on the inter- and intra-band scattering rates [7], [22]. Only if the anisotropies and diffusivities are nearly equal and interband scattering can be neglected, $H_{c2}(\theta)$ can be approximated by AGL in the whole $T$ range. However, it was shown earlier by Choi et al. [23] that AGL scaling of $J_c(\theta)$ for the two-band superconductor $MgB_2$ is possible.

Similar results on $LaFeAsO_{0.8}F_{0.2}$ thin films have been shown in [24]. Torque and resistive measurements on $SmFeAsO_{0.8}F_{0.2}$ and $NdFeAsO_{0.8}F_{0.2}$ revealed the same tem-perature dependence of $\gamma_H$ (in combination with an increasing anisotropy of the penetration depth $\gamma_\lambda$) [7]. K-doped Ba-122 shows the same $\gamma_H(T)$ dependence, however with slightly lower values [25]. This behavior of decreasing $H_{c2}$ anisotropy, $\gamma_H$, with decreasing $T$ corresponds to what has been found in dirty-limit $MgB_2$ thin films [22].

## V. Summary

We have deposited high-quality thin films of Co-doped Ba-122. The films grow epitaxially and are free of *c*-axis correlated defects. They show low $J_c$ values and small magnetic field anisotropies with the typical peak at $\theta = 90°$ for intrinsic pinning due to the layered crystal structure. No *c*-axis peaks ($\theta = 180°$) were observed. All $J_c(\theta)$ data of this multi-band superconductor were scaled using the anisotropic Ginzburg-Landau approach with one, yet temperature-dependent parameter, $\gamma_m$, for the electronic mass anisotropy. The anisotropy of the upper critical field, $\gamma_H$, determined resistively in pulsed-field measurements, matches the mass anisotropy obtained by the scaling approach. Both anisotropies, $\gamma_m$ and $\gamma_H$, are decreasing with decreasing temperature.

## Acknowledgment

The authors thank G. Fuchs and S.-L. Drechsler for fruitful discussions, J. Werner for the target preparation, as well as K. Nenkov, J. Engelmann, K. Tscharntke, M. Kühnel, C. Nacke, M. Deutschmann, and U. Besold for technical support.